# Mechanism of interaction of discrete and continuous variable states for efficient generation of hybrid entangled states

Sergey A. Podoshvedov

*Department of computer modeling and nanotechnology, Institute of natural and exact sciences, South Ural State University, Lenin Av. 76, Chelyabinsk, Russia*
e-mail: sapodo68@gmail.com

We develop theory of realization of hybrid entanglement between discrete-variable (single photon) and continuous-variable states (coherent states). A highly transmissive beam splitter (HTBS) is used for interaction of the ingredients. Mechanism for generating the hybrid entangled states is based on the simultaneous displacement of the discrete-variable state on equal in absolute value but opposite in sign displacement amplitudes by coherent components, in such a way that all the information about the displacement amplitudes is lost with subsequent registration of photons in the auxiliary mode. Conditions for the balanced hybrid entangled states generation are found. Our approach follows along lines experimentally demonstrated in Nat. Photonics 8, 570-574 (2014).

## 1. Introduction

In optical quantum information processing, there is two different approaches described in discrete-variable (DV) [1] and continuous-variable (CV) [2] framework. The methods use one of aspects of the particle-wave duality [3]. Many protocols employ the particle-like discrete nature of light, i.e. either the states $a_0|0\rangle + a_1|1\rangle$ or $a_0|H\rangle + a_1|V\rangle$, to encode quantum information. The states involve either vacuum and single photon or they are determined in polarization basis. Maximally entangled states from four-dimensional Hilbert space are required for implementing the protocols. In the continuous alternative, the state is determined in infinite dimensional space. For instance, the state $a_0|-\alpha\rangle + a_1|\alpha\rangle$ is superposition of classical light waves with opposite in sign amplitudes [4]. Both encodings have their inherent advantages and drawbacks [5-10]. The practical realization of Bell states detection being main ingredient of the quantum teleportation and construction of two-qubit operation controlled$-X$ [5] is major problem for their deterministic implementation since only two Bell states are discriminated by linear optics and photodetection [6], thus limiting the protocols to 50% [7,8] and 25% efficiency [9], respectively. Theoretical possibility in [1] can hardly be realized at least at the current level of technological development. In CV interpretation, the entangled resource corresponds to a two-mode squeezed state [2]. CV teleportation fidelity cannot reach unit as CV resources do not provide a maximally entangled state [10].

    Coherent states can be treated as macroscopic states of a light considered in CV framework. Single photon represents the best example of microscopic optical quantum system, which is usually described in the DV framework. A state that includes both DV and CV components can be called hybrid. Realization of the hybrid entanglement is of high interest for some reasons. On one side, it may answer fundamental questions underlying principles of the separation of the classical and quantum worlds, for example, represented by the so-called Schrödinger cat paradox [11] where the macroscopic classical states are entangled with microscopic quantum ones. On the other side, combining both approaches to realize hybrid architecture may overcome the current limitations [12-15]. Experimental aspects of the hybrid generation have been subject of intense research [16-18]. Different types of optical arrangements have been proposed for its generation. General features of the hybrid DC/CV entanglement that could be exploited for advanced quantum optical technologies are

analyzed in [19]. These proposals are based on various manifestations of the photon subtraction technique by recording a certain measurement outcome in an auxiliary mode.

Here, we study a more general mechanism for generating the hybrid entanglement using peculiarities of interaction of DV and CV states between each other (discrete-continuous interaction). This mechanism is based on indistinguishable displacement of DV state on equal in magnitude but opposite in sign amplitudes by components of superposition of coherent states (SCS) on HTBS [20]. The technique of the state displacement on the phase plane through the interaction of the original state with the coherent state of large amplitude has been successfully tested in [21,22]. The displacement of the state occurs in such a manner that all information about the amount by which the initial qubit is displaced is completely lost. Subsequent registration of photons in the auxiliary modes notifies the generation of the hybrid state. The feature of the discrete-continuous interaction on HTBS consists in the fact that the matrix amplitudes of the displaced states of light in the measurement number state basis change sign to the opposite when the displacement amplitude changes its sign [23-25]. It is related to action of the controlled $-Z$ gate on the DV qubit. This method is universal and allows us to realize different types of the hybrid states. In particular, the generation of two types of the hybrid states is considered. One of them consists of coherent components and single-rail qubit composed from vacuum and single photon. Another type is realized with dual-rail photon. The two schemes considered differ from each other only by an additional procedure of the displacement in the auxiliary mode. The proposed approach allows implementing both balanced with equal superposition weighs and unbalanced. The effectiveness of this method in terms of the success probability and the fidelity of the generated states are considered.

## 2. Discrete-continuous interaction on HTBS

Consider discrete-continuous-discrete interaction on HTBS which is described by the following unitary matrix

$$BS = \begin{bmatrix} t & -r \\ r & t \end{bmatrix}, \qquad (1)$$

where $t > 0, r > 0$ are the transmittance $t \to 1$ and reflectance $r \to 0$, respectively, satisfying the normalization condition $t^2 + r^2 = 1$. Before dealing with general case, consider interaction of two coherent states $|\alpha\rangle_1|\beta\rangle_2$ with amplitudes $\alpha$ and $\beta$, respectively, on HTBS, where the subscripts denote the modes. The interaction results in

$$BS|\alpha\rangle_1|\beta\rangle_2 = |\alpha t + \beta r\rangle_1|\beta t - \alpha r\rangle_2. \qquad (2)$$

Generalization of the relation for the arbitrary state $\rho$ gives [26]

$$BS(\rho \otimes |\beta\rangle\langle\beta|)BS^+ \approx D(\alpha)\rho D^+(\alpha) \otimes |\beta\rangle\langle\beta|, \qquad (3)$$

in the case of

$$\alpha = \beta r/t, \qquad (4)$$

where the displacement operator $D(\alpha)$ (A1) with displacement amplitude $\alpha$ is used, symbol $\otimes$ means tensor product of two operators and $D^+(\alpha)$ is Hermitian conjugate of the operator $D(\alpha)$. The same mathematical apparatus is applicable to interaction of arbitrary pure state with coherent state $|-\beta\rangle_2$ with output approximate state

$$BS(\rho \otimes |-\beta\rangle\langle -\beta|)BS^+ \approx D(-\alpha)\rho D^+(-\alpha) \otimes |-\beta\rangle\langle -\beta|, \qquad (5)$$

Note the condition (4) means that amplitude of coherent state may tend to infinity $\beta \to \infty$ if $r \to 0$ to keep the condition $\alpha = const$. In real experiment the non-zero reflectance $r \ne 0$ is used. The amplitude of the coherent states can take large but nevertheless finite values to satisfy the condition (4). For this reason, the sign of approximate equality is used in Eqs. (3) and (5) which goes into the exact equality in the limit case of $t \to 1$, $r \to 0$, and $\beta \to \infty$ to provide constant value of the displacement amplitude $\alpha$. The better we fulfill the condition $r \to 0$ and $\beta \to \infty$, the more accurately the output state will be to its ideal on the right-hand side of the Eqs. (3) and (5).

Now, consider interaction of one of SCS (for example even)
$$|even\rangle = N_+(|-\beta\rangle + |\beta\rangle), \tag{6a}$$
$$|odd\rangle = N_-(|-\beta\rangle - |\beta\rangle), \tag{6b}$$
where the factors $N_\pm = \left(2(1 \pm exp(-2|\beta|^2))\right)^{-1/2}$ are the normalization parameters, with dual-rail state of single photons superposition of two modes
$$|\varphi\rangle_{23} = a_0|01\rangle_{23} + a_1|10\rangle_{23}, \tag{7}$$
satisfying the normalization condition $|a_0|^2 + |a_1|^2 = 1$, on HTBS as shown in Fig. 1(a). In our case, single photon simultaneously takes modes 2 and 3. Here and in the following amplitude of SCS is assumed to be positive $\beta > 0$. Then, according to Eq. (4), the displacement amplitude take also positive values $\alpha > 0$. Then, using mathematical apparatus of Appendix 1, we have
$$BS_{13}(|even\rangle_1|\varphi\rangle_{23}) \to N_+ F \sum_{n=0}^\infty \frac{c_{1n}(\alpha)}{N_n N_n^{(tot)}} |\Psi_n\rangle_{12} |n\rangle_3, \tag{8}$$
in limit case of $t \to 1, r \to 0$. Here, we introduce the following two-mode entangled hybrid states
$$|\Psi_n\rangle_{12} = N_n^{(tot)} \left(|0,-\beta\rangle_1 |\varphi_n^{(+)}\rangle_2 + (-1)^{n-1} |0,\beta\rangle_1 |\varphi_n^{(-)}\rangle_2 \right), \tag{9}$$
with single-rail qubit composed from vacuum and single photon is
$$|\varphi_n^{(\pm)}\rangle_2 = N_n(a_0|0\rangle_2 \pm a_1 A_n |1\rangle_2). \tag{10}$$
Normalization factors of the states are defined by
$$N_n = (|a_0|^2 + |a_1|^2|A_n|^2)^{-1/2} = (1 + (|A_n|^2 - 1)|a_1|^2)^{-1/2}, \tag{11}$$
$$N_n^{(tot)} = \left(2(1 + (-1)^{n-1} N_n^2 exp(-2|\beta|^2)(1 - (1 + |A_n|^2)|a_1|^2))\right)^{-1/2}. \tag{12}$$
Note that other parameters $c_{1n}(\alpha)$ and $A_n$ are defined by Eqs. (A15) and (B4). The exponential factor $F$ is determined in Appendix A. Amplitude of the matrix element $\alpha$ is determined by Eq. (4). The designation (A4) for displaced number states is used in Eq. (9). The registration of $n$ photons in the auxiliary third mode makes it possible to generate the state (9). The success probability to produce the state $|\Psi_n\rangle_{12}$ in the limit case of is given by
$$P_n(\alpha) = N_+^2 F^2 \frac{|c_{1n}(\alpha)|^2}{N_n^2 N_n^{(tot)2}}. \tag{13}$$
The sum of the success probabilities is equal to one $\sum_{n=0}^\infty P_n(\alpha) = 1$ in the limit case of $t \to 1, r \to 0$ due to relation (A10).

Now, consider another type of interaction of continuous-variable state with entangled two-photon state occupying simultaneously four modes. For this purpose, the optical scheme in Fig. 1(b). It consists of two HTBS, where even SCS (6a) and additional coherent state $|0,-\beta_1\rangle_2$ ($\beta_1 > 0$) are mixed with two modes (modes 5 and 6, respectively) of two-photon entangled state
$$|\phi\rangle_{3456} = a_0|0101\rangle_{3456} + a_1|1010\rangle_{3456}, \tag{14}$$
Using mathematical approach presented in Appendix C, we have
$$BS_{15} BS_{26}(|even\rangle_1 |0,-\beta_1\rangle_2 |\phi\rangle_{3456}) \to$$
$$N_+ F^2 |0,-\beta\rangle_2 \sum_{n=0}^\infty \sum_{m=0}^\infty \frac{c_{0n}(\alpha) c_{1m}(\alpha_1)}{N_{nm} N_{nm}^{(tot)}} |\Psi_n\rangle_{134} |nm\rangle_{56}, \tag{15}$$
in limit case of $t \to 1$ provided that amplitudes $\alpha$ and $\alpha_1$ satisfy the condition (4) for input values of $\beta$ and $\beta_1$, respectively. Here, the following notations are introduced
$$|\Psi_{nm}\rangle_{134} = N_{nm}^{(tot)} \left(|0,-\beta\rangle_1 |\varphi_{nm}^{(+)}\rangle_{34} + (-1)^n |0,\beta\rangle_1 |\varphi_{nm}^{(-)}\rangle_{34} \right), \tag{16}$$
where dual-rail states of single photon are used
$$|\varphi_{nm}^{(\pm)}\rangle_{34} = N_{nm}(a_0|01\rangle_{34} \pm a_1 A_{nm}|10\rangle_{34}). \tag{17}$$
The normalization factors of the states are determined by

$$N_{nm} = (|a_0|^2 + |a_1|^2 |A_{nm}|^2)^{-1/2} = (1 + (|A_{nm}|^2 - 1)|a_1|^2)^{-1/2}, \quad (18)$$

$$N_{nm}^{(tot)} = \left(2\left(1 + (-1)^n N_{nm}^2 exp(-2|\beta|^2)(1 - (1 + |A_{nm}|^2)|a_1|^2)\right)\right)^{-1/2}. \quad (19)$$

Amplitude parameter $A_{nm}$ is presented by Eq. (C11). Measurement in the auxiliary modes 5 and 6 of $n$ and $m$ photons, respectively, leads to the generation of the hybrid entangled state (16) that comprises coherent states with opposite in signs amplitudes and dual-rail single photon. Success probability to generate the hybrid entangled state in the limit case of $t \to 1$, $r \to 0$ is

$$P_{nm}(\alpha, \alpha_1) = N_+^2 F^2(\alpha) F^2(\alpha_1) \frac{|c_{0n}(\alpha)|^2 |c_{1m}(\alpha_1)|^2}{N_{nm}^2 N_{nm}^{(tot)2}}, \quad (20)$$

Let us discuss physical basis for generation of the hybrid entangled states in Eqs. (9) and (16), respectively. Measurement of photons in the auxiliary modes leads to the generation of the hybrid entangled states. The key point to learn discrete-continuous interaction on HTBS is matrix elements of decomposition of one base set of the displaced number states (A5) over another [23]. In our case, we are interested in the decomposition of the displaced number states over number ones (A6). The matrix elements (A8) are written as a product of $\alpha^{n-l}$ and another factor bracketed. The bracketed factor is some polynomial in absolute values of the displacement amplitude $|\alpha|$. It is natural the polynomial does not affect the shape of the matrix elements when we rotate the displacement amplitude in the phase plane by some angle $\alpha \to \alpha exp(i\varphi)$. Conversely, the factor $\alpha^{n-l}$ determines the phase of the matrix elements when the displacement amplitude is rotated in the phase plane. So, the matrix elements of the zero row of the transformation matrix $c_{0n}(\alpha)$ in Eq. (A14) are the elements of the coherent state. They change as $(-1)^n$ under change of the displacement amplitude on opposite $\alpha \to -\alpha$. The matrix element of the first row of the same transformation matrix $c_{1n}(\alpha)$ in Eq. (A15) determines the displaced single photon. The matrix elements change as $(-1)^{n-1}$ under change of the displacement amplitude on opposite. This difference in factors when the parity of the displaced state is changed is akin to nonlinear action of two-qubit controlled $Z-$ gate. Note that the mechanism does not work for the target entangled states in the polarization encoding like

$$|\varphi\rangle_2 = a_0 |H\rangle_{23} + a_1 |V\rangle_2, \quad (21)$$

or

$$|\phi\rangle_{34} = a_0 |HV\rangle_{34} + a_1 |VH\rangle_{34}, \quad (22)$$

as the states $|H\rangle$ and $|V\rangle$ have the same parity. The coherent components of the SCS (6a) simultaneously displace the target qubit in indistinguishable manner on HTBS by the values that differ from each other only by sign. All information about the values of the displacement of the target qubit disappears and this uncertainty forms superposition hybrid state. Additional amplitude factors $A_n$ and $A_{nm}$ follows from the matrix elements and they are inherent part of the discrete-continuous interaction on HTBS. A coherent auxiliary state $|0, -\beta\rangle_2$ is used to realize the state in Eq. (16). The state does not change the sign of the matrix elements $c_{0n}(\alpha)$ and $c_{1n}(\alpha)$.

Here, we can estimate how the ideal generated state (9) is close to real in the case of $t \neq 1$ and $r \neq 0$ obtained by measurement procedure. Using expressions (B16), it is possible to estimate the fidelities of the generated states

$$Fid_1 = \left(N_n^{(tot)}(\beta) N_n^{(tot)}(\beta/t) N_n^{(tot)'}(\beta)\right)^2, \quad (23)$$

where

$$N_n^{(tot)'}(\beta) = 2 \left( \frac{exp\left(-\frac{|\beta|^2}{2}\left(1-\frac{1}{t}\right)^2\right) + (-1)^{n-1} N_n^2 exp\left(-\frac{|\beta|^2}{2}\left(1+\frac{1}{t}\right)^2\right)}{(1 - (1 + |A_n|^2)|a_1|^2)} \right). \quad (24)$$

Doing the same calculations, we can estimate fidelity for the state (16) to be generated in zero order in the parameter $r \ll 1$

$$Fid_2 = \left(N_{nm}^{(tot)}(\beta)N_{nm}^{(tot)}(\beta/t)N_{nm}^{(tot)'}(\beta)\right)^2, \qquad (25)$$

where the quantities $N_{nm}^{(tot)}(\beta)$ and $N_{nm}^{(tot)}(\beta/t)$ are given by Eq. (19) while $N_{nm}^{(tot)'}(\beta)$ follows from Eq. (24) by substitution of $A_n \to A_{nm}$ and $N_n \to N_{nm}$. In the general case, the fidelities depend in a complex manner both on the parameters of the experimental setup, the amplitude of the coherent components, displacement amplitude $\alpha$, and on the value of the absolute value $|a_1|$ of the target states in Eqs. (7) and (14).

## 3. Generation of the hybrid entangled states

The previous part is the basis for considering generation of the hybrid states. If $n$ photons is registered at auxiliary third mode in Fig. 1(a), then the state (9) with fidelity determined by Eq. (24) is produced [17]. In the general case, this state is unbalanced in the sense that its amplitudes are not equal in absolute value to each other $|a_0| \neq |a_1||A_n|$. Nevertheless, we can consider the conditions under which balanced hybrid entangled state

$$\left|\Psi_n^{(b)}\right\rangle_{12} = \frac{1}{\sqrt{2}}\begin{pmatrix} |0,-\beta\rangle_1 \frac{1}{\sqrt{2}}(|0\rangle_2 + exp(i\delta)|1\rangle_2) + \\ (-1)^{n-1}|0,\beta\rangle_1(|0\rangle_2 - exp(i\delta)|1\rangle_2) \end{pmatrix}, \qquad (26)$$

is produced, where the condition $|a_0| = |a_1||A_n|$ is performed and $\delta$ is the relative phase of amplitudes $a_0$ and $a_1$. Here, the superscript $(b)$ means the balanced state. Applying the Hadamard transformation $H = |0\rangle(\langle 0|+\langle 1|)/\sqrt{2} + |1\rangle(\langle 0|-\langle 1|)/\sqrt{2}$, one obtains

$$\left|\Psi_n^{(b)}\right\rangle_{12} = \frac{1}{\sqrt{2}}(|0,-\beta\rangle_1|0\rangle_2 + (-1)^{n-1}|0,\beta\rangle_1|1\rangle_2), \qquad (27)$$

where $\delta = 0$ is taken. The condition of generation of the balanced hybrid entangled state (27) is provided by $|a_0|^2 = |A_n|^2/(1+|A_n|^2)$ and $|a_1|^2 = 1/(1+|A_n|^2)$ that leads to success probability of the event

$$P_n^{(b)}(\alpha) = 4N_+^2 F^2 |c_{1n}(\alpha)|^2 \frac{|A_n(\alpha)|^2}{1+|A_n(\alpha)|^2}, \qquad (28)$$

with fidelity in the zeroth order of parameter $r \ll 1$

$$Fid_1 = exp\left(-|\alpha|^2 \frac{1-t}{1+t}\right). \qquad (29)$$

In this case, success probability and fidelity do not depend on the amplitude $|a_1|$ of the auxiliary qubits. Corresponding plots in Figs. 2(a-c) show dependence of the success probabilities $P_n^{(b)}(\alpha)$ for $n = 0$ (Fig. 2(a)) and $n = 1$ (Fig. 2(b)) and their fidelities (Fig. 2(c)) being equal to each other on the displacement amplitude $\alpha$. The curves are obtained for different values of the transmittance $t$. An increase in the value of this parameter makes it possible to some extent to improve the success probabilities to produce balanced hybrid entangled state (27). The fidelity of the generated states can also be increased with an increase of the transmittance $t$ (Fig. 2(c)). Using the values of the displacement amplitude, we can get the value of amplitudes $\beta$ of the coherent components of the hybrid state by using the expression (4).

Consider the possibility of generation of the hybrid entangled state that contains coherent components and dual-rail single photon (16). The states are produced by simultaneous registration of the states $|n\rangle_5$ and $|m\rangle_6$ in the fifth and sixth auxiliary modes, respectively, as shown in Fig. 1(b). In this case it is possible as the generation of unbalanced states (16) with $|a_0| \neq |a_1||A_{nm}|$ and balanced ones

$$\left|\Psi_{nn}^{(b)}\right\rangle_{123} = \frac{1}{\sqrt{2}}\begin{pmatrix} |0,-\beta\rangle_1 \frac{1}{\sqrt{2}}(|01\rangle_{23} + exp(i\delta)|10\rangle_{23}) + \\ (-1)^n|0,\beta\rangle_1 \frac{1}{\sqrt{2}}(|01\rangle_{23} - exp(i\delta)|10\rangle_{23}) \end{pmatrix}, \qquad (30)$$

satisfying the condition $|a_0| = |a_1| = 1/\sqrt{2}$ and $|A_{nm}| = 1$. The state (30) can be transformed into

$$|\Psi_{nn}^{(b)}\rangle_{123} = \tfrac{1}{\sqrt{2}}(|0,-\beta\rangle_1|01\rangle_{23} + (-1)^n|0,\beta\rangle_1|10\rangle_{23}), \tag{31}$$

by applying the Hadamard transformation to the dual-rail single photon in the case of $\delta = 0$. We note that the Hadamard transform on the dual-rail single photon can be easily carried out by the methods of linear optics [27] unlike single-rail qubit. Here the superscript $(b)$ denotes the balanced state. It follows from Eq. (C11), the condition $|A_{nm}| = 1$ is performed in the case of $\alpha = \alpha'$ and $n = m$. The success probability to realize the balanced hybrid entangled state (30) follows from Eq. (20)

$$P_{nm}^{(b)}(\alpha) = 2N_+^2 F^4(\alpha)|c_{0n}(\alpha)|^2|c_{1n}(\alpha)|^2. \tag{32}$$

The fidelity to generate the hybrid entangled state (30) coincides with (29) derived for production of the state (26). If the parities of the detected photons in the auxiliary fifth and sixth modes in Fig. 1(b) are different from each other $n \neq m$, then some terms of the superposition state

$$|\Psi_{nm}\rangle_{123} = N_{nm}^{(tot)}\begin{pmatrix} |0,-\beta\rangle_1 N_{nm}(|01\rangle_{23} + exp(i\delta)A_{nm}|10\rangle_{23})/\sqrt{2} + \\ (-1)^n|0,\beta\rangle_1 N_{nm}(|01\rangle_{23} - exp(i\delta)A_{nm}|10\rangle_{23})/\sqrt{2} \end{pmatrix} \tag{33}$$

acquires an additional amplitude factor $A_{nm}$ in the case of $|a_0| = |a_1| = 1/\sqrt{2}$. Corresponding dependencies of $P_{00}(\alpha)$ and $P_{11}(\alpha)$ to generate balanced hybrid superposition on the displacement amplitude $\alpha$ are shown in Figs. 3(a,b), respectively, for different values of the experimental parameter $t$. As can be seen from the figures, an increase in the value of the parameter $t$ leads to a slight increase in the success probability under small values of the displacement amplitude $\alpha$. Plot in figure 3(c) shows dependence of the success probabilities $P_{01}(\alpha) = P_{10}(\alpha)$ to generate the unbalanced hybrid entangled state (33) on the displacement amplitude $\alpha$. The probability of success decreases with increasing parameter $\alpha$. Note that the state (31) can be transformed into the hybrid entangled state with single photon in polarization basis

$$|\Psi_{nn}^{(b)}\rangle_{12} = \tfrac{1}{\sqrt{2}}(|0,-\beta\rangle_1|H\rangle_2 + (-1)^n|0,\beta\rangle_1|V\rangle_2), \tag{34}$$

with help of polarizer and polarization beam splitter. In practice, implementing the Hadamard operation for a single-rail qubit by linear optics is a rather complicated problem. But we can realize the state (27) by the same technique that was used to generate the states (9) and (16). Mixing mode three of the state (31) with a coherent state with amplitude $\gamma$ such that the condition $c_{0n}(\gamma) = c_{1n}(\gamma)$ is performed and registering $n$ photons in the auxiliary mode, one obtains the state (27). This condition is satisfied in the case of the displacement amplitude $\gamma_1 = (-1 + \sqrt{1+4n})/2$ and $\gamma_2 = (-1 - \sqrt{1+4n})/2$.

Consider opportunities to increase the success probability to generate the balanced hybrid entangled state (30). So, we can impose the condition $A_{nm} = -1$ in the case of $n \neq m$ which can be realized for $|\alpha|^2 = (n+m)/2$. For example, consider the cases either $n = 0, m = 1$ or $n = 1, m = 0$ that gives $|\alpha| = 1/\sqrt{2}$. Use of this value of the displacement amplitude makes it possible to additionally generate the balanced hybrid entangled state (30) in the case of registration of the states either $|01\rangle_{56}$ or $|10\rangle_{56}$. Thus, the state (30) is generated both when recording photons with the same parity in auxiliary modes and when detecting single photon in one mode and vacuum in another for the displacement amplitude. Corresponding plot in Fig. 4 shows dependence of the success probability to realize hybrid entangled state (30) on the transmittance $t$ in the case of $|\alpha| = 1/\sqrt{2}$. Here, four measurement outcomes $|00\rangle_{56}$, $|01\rangle_{56}$, $|10\rangle_{56}$ and $|11\rangle_{56}$ give their contribution to the generation. An increase in the probability of success compared to the figures 3(a) and 3(b) is observed.

There is another possibility to produce the state by imposing the condition $|a_0| = |a_1||A_{nm}|$ that is valid in the case of $|a_1|^2 = 1/(1 + |A_{nm}|^2)$. Then, when registering $n$ photons in fifth mode and $m$ photons in sixth mode, what happens with the probability of success

$$P_{nm}^{(b)}(\alpha) = 4N_+^2 F^4 |c_{0n}(\alpha)|^2 |c_{1n}(\alpha)|^2 \frac{|A_{nm}(\alpha)|^2}{1+|A_{nm}(\alpha)|^2}, \tag{35}$$

we generate the balanced hybrid entangled state (30) with the fidelity given by Eq. (29). Numerical results show similar dependencies of the success probability on the displacement amplitude as in figures 3.

In the original work [17], the possibility of generating the hybrid state with coherent and single-rail components was experimentally demonstrated on example of the target two-mode squeezed state

$$|\varphi_1\rangle_{si} = N_\lambda (|00\rangle_{si} + \lambda |11\rangle_{si}), \tag{36}$$

which is generated on output from continuous-wave optical parametric oscillator for the signal $s$ and idler $i$ modes in the very low gain limit, where $N_\lambda$ is normalization factor. Then, using the same calculation technique for Fig. 1(a), we can obtain the state (9) with single-rail components

$$\left|\varphi_n^{(\pm)}\right\rangle_2 = N_n (|0\rangle_2 \pm \lambda A_n^{-1} |1\rangle_2). \tag{37}$$

The success probability of the generation of the state is the same as given by the expression (13) taking into account the replacement $c_{1n}(\alpha) \to c_{0n}(\alpha)$ and $A_{nm} \to A_{nm}^{-1}$. Imposition of the condition $\lambda A_n^{-1} = 1$ gives a possibility to produce the balanced hybrid entangled state (26).

Here, we also give an example of hybrid converter enabling to map discrete qubits to coherent state qubits. Consider hybrid entangled state (31) and use beam splitter in Eq. (1) to transform the single photon as $|01\rangle_{23} \to r|10\rangle_{23} + t|01\rangle_{23}$ and $|10\rangle_{12} \to t|10\rangle_{12} - r|01\rangle_{12}$. Then, due to simultaneous presence of the single photon in both modes and indistinguishability between the events, the following superposition states are expected to be generated

$$|\Phi_1\rangle_1 = N(t|-\beta\rangle - r|\beta\rangle), \tag{38a}$$
$$|\Phi_2\rangle_1 = N(r|-\beta\rangle + r|\beta\rangle), \tag{38b}$$

provided that the following measurement outcomes $|01\rangle_{23}$ and $|10\rangle_{23}$, respectively, are registered. Here, $N$ is the corresponding normalization factor. The states (38a) and (38b) can be considered as qubits subjected to a one-qubit transformation in basis $|\pm \alpha\rangle$.

## 4. Results

We developed the theory of interaction of the discrete-variable and continuous-variable (discrete-continuous interaction) states between each other on HTBS. Generation of the hybrid entanglement between discrete-variable states (single photons) and continuous-variable states (coherent states) has special interest in fundamental quantum mechanics within the framework of demonstration of Schrödinger cat states [11]. The theory is based on peculiarities of the discrete-continuous interaction on HTBS. Manipulations with DV and CV states are well known, while the discrete-continuous interaction is a fairly new and unknown problem, the solution of which can give serious advantages for the implementation of quantum protocols [15]. Coherent components of the original SCS displace the target microscopic state on different on sign quantities in an indistinguishable manner. The subsequent registration of photons in the auxiliary mode generates a set of the hybrid entangled states whose amplitudes differ from each other by known values. Change in the sign of the displacement amplitude to the opposite leads to the fact that the amplitudes of the decomposition of the displaced states of light vary depending on the overall parity of the photon being registered and the target displaced state. Thus, the hybrid entanglement arises irrespective of the number of registered photons due to action of the mechanism of discrete-continuous interaction. In general case, the generated hybrid states only have different amplitudes but the initial conditions can be chosen so as to equalize the amplitudes of the hybrid state.

Two optical schemes to realize two different types of the hybrid states are considered. One of them is formed from coherent components and single-rail qubit composed of vacuum and single photon. Either single photon occupying two modes (7) or two-mode squeezed state with small value of the squeezing parameter (36) can act as the target state for generation of the hybrid states. Another type of hybrid entanglement (16) can be obtained by using two-photon state in mode representation (14). The state (14) can be obtained by converting input polarization entangled state (22) by linear optics tools. In this case, this state becomes ready for application of the mechanism considered. Even more, by generating the hybrid states with dual-rail photon as component, one can also return to the original polarization basis (34). An additional mixing of the target state with a coherent state with subsequent recording of measurement outcomes by two detectors is required to generate the output states. The superposition weighs can be balanced under certain conditions found in the work. The success probabilities to generate the states of two types in dependence on the displacement amplitude and the HTBS parameter are received. The best strategies of generation of the balanced hybrid states with the greatest success probabilities and maximum fidelity are considered. The schemes rely on a probabilistic preparation heralded by the detection of photons. In this way, noisy environment affects count rate but not the fidelity of the resulting states. Therefore, the proposed method is more suitable to establish entanglement between different subsystems that differ in size or in the way they are most conveniently described in discrete- and continuous-variable framework. Note that the implementation of these optical circuits requires a minimum irreducible number of optical elements, which increases the effectiveness of this approach.

**Appendix A. Properties of matrix elements of displacement operator**

Next consideration is based on use of unitary displacement operator determined by
$$D(\alpha) = exp(\alpha a^+ - \alpha^* a), \quad (A1)$$
where $\alpha$ is an amplitude of the displacement and $a$, $a^+$ are the bosonic annihilation and creation operators. Its action on arbitrary pure state
$$|\psi\rangle = \sum_{n=0}^{k} a_n |n\rangle, \quad (A2)$$
where the normalization condition for the superposition $\sum_{n=0}^{k} |a_n|^2$ holds, is the following
$$D(\alpha)|\psi\rangle = D(\alpha) \sum_{n=0}^{k} a_n |n\rangle = \sum_{n=0}^{k} a_n |n, \alpha\rangle, \quad (A3)$$
where the notation for the displaced number states
$$|n, \alpha\rangle = D(\alpha)|n\rangle, \quad (A4)$$
is used [23]. The displaced number states (A4) are defined by two numbers: quantum discrete number $n$ and classical continuous parameter $\alpha$ which can be recognized as their size [23]. The states (A4) belong to vector (infinite Hilbert) space with appropriate inner product $\langle n, \alpha | m, \alpha \rangle = \delta_{nm}$ with $\delta_{nm}$ being Kronecker delta [5], which is determined by the displacement amplitude $\alpha$ with the base states
$$\{|n, \alpha\rangle, n = 0,1,2, \ldots, \infty\}. \quad (A5)$$
The basic set of displaced number states for an arbitrary displacement parameter $\alpha$ is complete, which allows us to decompose an arbitrary state over the base states. Consider the decomposition on example of displaced number states [20]
$$|l, \alpha\rangle = F \sum_{n=0}^{\infty} c_{ln}(\alpha) |n\rangle, \quad (A6)$$
where multiplier $F = exp(-|\alpha|^2/2)$ is introduced and the matrix elements are the following [23]
$$c_{ln}(\alpha) = \frac{\alpha^{n-l}}{\sqrt{l!n!}} \sum_{k=0}^{l} (-1)^k C_l^k |\alpha|^{2k} \prod_{k=0}^{l-1} (n - l + k + 1), \quad (A7)$$
or the same

$$c_{ln}(\alpha) = \frac{\alpha^{n-l}}{\sqrt{l!n!}} \begin{pmatrix} n(n-1)\ldots(n-l+1) - C_l^1|\alpha|^2 n(n-1)\ldots(n-l+2) + \\ C_l^2|\alpha|^4 n(n-1)\ldots(n-l+3) + \cdots \\ (-1)^k C_l^k |\alpha|^{2k} \prod_k^{l-1}(n-l+k+1) + (-1)^l |\alpha|^{2l} \end{pmatrix}, \quad (A8)$$

where $C_l^k = l!/(k!(l-k)!)$ are the elements of the Bernoulli distribution and number product is

$$\prod_k^{l-1}(n-l+k+1) = n(n-1)\ldots(n-l+k+1). \quad (A9)$$

It is possible directly to check the matrix elements $c_{mn}(\alpha)$ (A7) satisfy the normalization condition [23]

$$F^2 \sum_{n=0}^{\infty} |c_{ln}(\alpha)|^2. \quad (A10)$$

It is worth noting that the reverse transformation $c_{ln}(-\alpha)$ defines number state through their displaced analogies due to unitary nature of the displacement operator (A1).

Let us analyze structure of the matrix elements (A7). The matrix elements consist of a common factor proportional to $\alpha^{n-l}$ and a polynomial expression of degree $l$ being the maximum of the degree of its monomial over $|\alpha|^2$ which is enclosed in parentheses. The polynomial in parentheses is invariant when changing variables $\to \alpha exp(i\phi)$, where $\phi$ is an arbitrary phase. The term in front of the polynomial $\alpha^{n-l}$ defines the behavior of the matrix elements under change $\alpha \to \alpha exp(i\phi)$. In particular, it determines behavior of the matrix elements under change the displacement amplitude on opposite $\alpha \to -\alpha$ that leads to

$$c_{ln}(-\alpha) = (-1)^{n-l} c_{ln}(\alpha). \quad (A11)$$

In particular, we have the following rules for decomposition of the even displaced number states $l = 2m$

$$c_{2mn}(-\alpha) = (-1)^n c_{2mn}(\alpha), \quad (A12)$$

and of the odd displaced number states $l = 2m + 1$

$$c_{2m+1n}(-\alpha) = (-1)^{n-1} c_{2m+1n}. \quad (A13)$$

Consider partial cases with $l = 0$ and $l = 1$ corresponding to the decomposition of coherent state and displaced single photon over the number states [23], whose matrix elements are the following

$$c_{0n}(\alpha) = \frac{\alpha^n}{\sqrt{n!}}, \quad (A14)$$

$$c_{1n}(\alpha) = \frac{\alpha^{n-1}}{\sqrt{n!}}(n - |\alpha|^2), \quad (A15)$$

satisfying the normalization conditions

$$F^2 \sum_{n=0}^{\infty} |c_{0n}(\alpha)|^2 = 1, \quad (A16)$$
$$F^2 \sum_{n=0}^{\infty} |c_{1n}(\alpha)|^2 = 1. \quad (A17)$$

In application to the matrix elements of coherent and displaced single photon states, we have

$$c_{0n}(-\alpha) = (-1)^n c_{0n}(\alpha), \quad (A18)$$
$$c_{1n}(-\alpha) = (-1)^{n-1} c_{1n}(\alpha). \quad (A19)$$

**Appendix B. Peculiarities of discrete-continuous interaction of even SCS with dual-rail single photon on HTBS**

Let us analyze in more detail interaction of even SCS (6a) with dual-rail single photon (7) on HTBS (1). Due to linearity of the beam splitter operator, we have

$$BS_{13}(|even\rangle_1 |\varphi\rangle_{23}) = N_+ \big(BS_{13}(|0, -\beta\rangle_1 |\varphi\rangle_{23}) + BS_{13}(|0, \beta\rangle_1 |\varphi\rangle_{23})\big). \quad (B1)$$

Consider action of the beam splitter on the states separately. Then, we have

$$BS_{13}(|0, -\beta\rangle_1 |\varphi\rangle_{23}) = BS_{13} D_1(-\beta) D_3(-\alpha) BS_{13}^+ BS_{13} |0\rangle_1 D_3(\alpha) |\varphi\rangle_{23} =$$
$$D_1(-\beta/t) D_3(0) BS_{13} |0\rangle_1 (a_0 |0\rangle_2 |1, \alpha\rangle_3 + a_1 |1\rangle_2 |0, \alpha\rangle_3) =$$
$$FD_1(-\beta/t) \sum_{n=0}^{\infty} c_{1n}(\alpha) (a_0 |0\rangle_2 + a_1 A_n |1\rangle_2) BS_{13}(|0n\rangle_{13}). \quad (B2)$$

The same mathematical apparatus is applicable to the second term

$$BS_{13}(|0,\beta\rangle_1|\varphi\rangle_{23}) = BS_{13}D_1(\beta)D_3(\alpha)BS_{13}^+BS_{13}|0\rangle_1 D_3(-\alpha)|\varphi\rangle_{23} =$$
$$D_1(\beta/t)D_3(0)BS_{13}|0\rangle_1(a_0|0\rangle_2|1,-\alpha\rangle_3 + a_1|1\rangle_2|0,-\alpha\rangle_3) =$$
$$FD_1(-\beta/t)\sum_{n=0}^{\infty}(-1)^{n-1}c_{1n}(\alpha)(a_0|0\rangle_2 - a_1 A_n|1\rangle_2)BS_{13}(|0n\rangle_{13}). \quad (B3)$$

Here, we have used the expansion (A6) and the dependence of the matrix coefficients as a function of the number $n$ when replacing $\alpha \to -\alpha$ in Eqs. (A11, A18, A19). Amplitude parameter $A_n$ is defined by

$$A_n(\alpha) = \frac{c_{0n}(\alpha)}{c_{1n}(\alpha)} = \frac{\alpha}{n-|\alpha|^2}. \quad (B4)$$

Summing up the expressions, we have

$$BS_{13}(|even\rangle_1|\varphi\rangle_{23}) = N_+ F$$
$$\begin{pmatrix} |\Delta_0(-\beta)\rangle_{123} + |\Delta_0(\beta)\rangle_{123} + r(|\Delta_1(-\beta)\rangle_{123} + |\Delta_1(\beta)\rangle_{123}) + \\ r^2(|\Delta_2(-\beta)\rangle_{123} + |\Delta_2(\beta)\rangle_{123}) + \cdots \end{pmatrix}, \quad (B5)$$

where

$$|\Delta_0(-\beta)\rangle_{123} = |0,-\beta/t\rangle_1 \sum_{n=0}^{\infty} c_{1n}(\alpha) t^n (a_0|0\rangle_2 + a_1 A_n|1\rangle_2)|n\rangle_3, \quad (B6)$$

$$|\Delta_0(\beta)\rangle_{123} = |0,\beta/t\rangle_1 \sum_{n=0}^{\infty}(-1)^{n-1} c_{1n}(\alpha) t^n (a_0|0\rangle_2 - a_1 A_n|1\rangle_2)|n\rangle_3, \quad (B7)$$

$$|\Delta_1(-\beta)\rangle_{123} = |1,-\beta/t\rangle_1 \sum_{n=1}^{\infty} c_{1n}(\alpha) t^{n-1}\sqrt{n}(a_0|0\rangle_2 + a_1 A_n|1\rangle_2)|n-1\rangle_3, \quad (B8)$$

$$|\Delta_1(\beta)\rangle_{123} = |1,\beta/t\rangle_1 \sum_{n=1}^{\infty}(-1)^{n-1} c_{1n}(\alpha) t^{n-1}\sqrt{n}(a_0|0\rangle_2 - a_1 A_n|1\rangle_2)|n-1\rangle_3, (B9)$$

$$|\Delta_2(-\beta)\rangle_{123} = |2,-\beta/t\rangle_1 \sum_{n=2}^{\infty} c_{1n}(\alpha) t^{n-2}\sqrt{\frac{n(n-1)}{2!}}(a_0|0\rangle_2 + a_1 A_n|1\rangle_2)|n-2\rangle_3, (B10)$$

$$|\Delta_2(\beta)\rangle_{123} = |2,\beta/t\rangle_1$$
$$\sum_{n=2}^{\infty}(-1)^{n-1} c_{1n}(\alpha) t^{n-2}\sqrt{\frac{n(n-1)}{2!}}(a_0|0\rangle_2 - a_1 A_n|1\rangle_2)|n-2\rangle_3. \quad (B11)$$

Using the expressions, we can obtain the formulas (8-12) in the case of $t \to 1$.

Let us now present derivation of analytical expression for the fidelity which shows how close the two states are to each other. Suppose $|n\rangle$ photon state is registered at third auxiliary mode. Then, we have the following state

$$|\Psi_r^{(n)}\rangle_{12} = N_r^{(n)}\left(|\Psi_0^{(n)}\rangle_{12} + r\sqrt{n+1}\frac{c_{1n+1}}{c_{1n}}|\Psi_1^{(n)}\rangle_{12}\right), \quad (B12)$$

taking into account only the terms of the first order of smallness on parameter $r \ll 1$. Here, the following states are introduced

$$|\Psi_0^{(n)}\rangle_{12} = N_n^{(tot)}(\beta/t)\left(|0,-\beta/t\rangle_1|\varphi_n^{(+)}\rangle_2 + (-1)^{n-1}|0,\beta/t\rangle_1|\varphi_n^{(-)}\rangle_2\right), \quad (B13)$$

$$|\Psi_1^{(n)}\rangle_{12} = N_{n1}^{(tot)}(\beta/t)\left(|1,-\beta/t\rangle_1|\varphi_{n+1}^{(+)}\rangle_2 + (-1)^n|1,\beta/t\rangle_1|\varphi_{n+1}^{(-)}\rangle_2\right), \quad (B14)$$

with corresponding normalization factors $N_n^{(tot)}$ (Eq. (12)) and $N_{n1}^{(tot)}$, respectively. The common normalization factor $N_r^{(n)}$ is given by

$$N_r^{(n)} = \begin{pmatrix} 1 + r^2(n+1)\frac{|c_{1n+1}|^2}{|c_{1n}|^2} + \\ r\sqrt{n+1}\left(\frac{c_{1n+1}}{c_{1n}}\langle\Psi_0^{(n)}|\Psi_1^{(n)}\rangle + \left(\frac{c_{1n+1}}{c_{1n}}\right)^*\langle\Psi_1^{(n)}|\Psi_0^{(n)}\rangle\right) \end{pmatrix}^{-1/2}. \quad (B15)$$

Ideal state follows from Eq. (9). Then, the fidelity between the states can be derived as

$$Fid_1 = |\langle\Psi_n|\Psi_r^{(n)}\rangle|^2 = N_r^{(n)2}$$
$$\begin{pmatrix} |\langle\Psi_n|\Psi_0^{(n)}\rangle|^2 + r\sqrt{n+1} \\ \left(\left(\frac{c_{1n+1}}{c_{1n}}\right)^*\langle\Psi_n|\Psi_0^{(n)}\rangle\langle\Psi_n|\Psi_1^{(n)}\rangle^* + \frac{c_{1n+1}}{c_{1n}}\langle\Psi_n|\Psi_0^{(n)}\rangle^*\langle\Psi_n|\Psi_1^{(n)}\rangle\right) \end{pmatrix}, \quad (B16)$$

up to the first order in the parameter $r \ll 1$. Using the expression, we can estimate the fidelity by (23).

**Appendix C. Peculiarities of discrete-continuous interaction of even SCS with two-photon entangled state on HTBS**

Now, we are going to make use of the same mathematical apparatus as in Appendix B to derive expression (15). We have the following

$$BS_{15}BS_{26}(|even\rangle_1|0,-\beta_1\rangle_2|\phi\rangle_{3456}) =$$
$$N_+\big(BS_{15}BS_{26}(|0,-\beta\rangle_1|0,-\beta_1\rangle_2|\phi\rangle_{3456}) + BS_{15}BS_{26}(|0,\beta\rangle_1|0,-\beta_1\rangle_2|\phi\rangle_{3456})\big). \quad (C1)$$

As well as in an Appendix B examine each term separately. Then, one obtains for first term

$$BS_{15}BS_{26}(|0,-\beta\rangle_1|0,-\beta_1\rangle_2|\phi\rangle_{3456}) =$$
$$BS_{15}D_1(-\beta)D_5(-\alpha)BS_{15}^+BS_{26}D_2(-\beta_1)D_6(-\alpha_1)BS_{26}^+BS_{15}BS_{26}|00\rangle_{12}D_5(\alpha)D_6(\alpha_1)|\phi\rangle_{3456} =$$
$$D_1(-\beta/t)D_5(0)D_2(-\beta_1/t)D_6(0)BS_{15}BS_{26}|00\rangle_{12}(a_0|01\rangle_{34}|0,\alpha\rangle_5|1,\alpha_1\rangle_6 +$$
$$a_1|10\rangle_{34}|1,\alpha\rangle_5|0,\alpha_1\rangle_6), \quad (C2)$$

Applying the same technique, we get for the second term

$$BS_{15}BS_{26}(|0,\beta\rangle_1|0,-\beta_1\rangle_2|\phi\rangle_{3456}) =$$
$$BS_{15}D_1(\beta)D_5(\alpha)BS_{15}^+BS_{26}D_2(-\beta_1)D_6(-\alpha_1)BS_{26}^+BS_{15}BS_{26}|00\rangle_{12}D_5(-\alpha)D_6(\alpha_1)|\phi\rangle_{3456} =$$
$$D_1(\beta/t)D_5(0)D_2(-\beta_1/t)D_6(0)BS_{15}BS_{26}|00\rangle_{12}(a_0|01\rangle_{34}|0,-\alpha\rangle_5|1,\alpha_1\rangle_6 +$$
$$a_1|10\rangle_{34}|1,-\alpha\rangle_5|0,\alpha_1\rangle_6). \quad (C3)$$

Let us embrace the decomposition (A6) together with Eqs. (A18) and (A19), to finally obtain

$$BS_{15}BS_{26}(|even\rangle_1|0,-\beta_1\rangle_2|\phi\rangle_{3456}) = N_+F^2, \quad (C4)$$

$$\left(\begin{array}{c}|\Delta_0(-\beta,-\beta_1)\rangle_{123456} + |\Delta_0(\beta,-\beta_1)\rangle_{123456} + r(|\Delta_1(-\beta,-\beta_1)\rangle_{123456} + |\Delta_1(\beta,-\beta_1)\rangle_{123456}) + \\ r^2(|\Delta_2(-\beta)\rangle_{123} + |\Delta_2(\beta)\rangle_{123}) + \cdots\end{array}\right)$$

where the following states are introduced

$$|\Delta_0(-\beta,-\beta_1)\rangle_{123456} = |0,-\beta/t\rangle_1|0,-\beta_1/t\rangle_2$$
$$\sum_{n=0}^{\infty}\sum_{m=0}^{\infty} c_{0n}(\alpha)c_{1m}(\alpha_1)(a_0|01\rangle_{34} + a_1A_{nm}|10\rangle_{34})|nm\rangle_{56}, \quad (C5)$$

$$|\Delta_0(\beta,-\beta_1)\rangle_{123456} = |0,\beta/t\rangle_1|0,-\beta_1/t\rangle_2$$
$$\sum_{n=0}^{\infty}\sum_{m=0}^{\infty} (-1)^n c_{0n}(\alpha)c_{1m}(\alpha_1)(a_0|01\rangle_{34} - a_1A_{nm}|10\rangle_{34})|nm\rangle_{56}, \quad (C6)$$

$$|\Delta_1(-\beta,-\beta_1)\rangle_{123456} = |1,-\beta/t\rangle_1|0,-\beta_1/t\rangle_2$$
$$\sum_{n=1}^{\infty}\sum_{m=0}^{\infty} c_{0n}(\alpha)c_{1m}(\alpha_1)\sqrt{n}t^{n+m-1}(a_0|01\rangle_{34} + a_1A_{nm}|10\rangle_{34})|n-1m\rangle_{56} +$$
$$|0,-\beta/t\rangle_1|1,-\beta_1/t\rangle_2 \sum_{n=0}^{\infty}\sum_{m=1}^{\infty} c_{0n}(\alpha)c_{1m}(\alpha_1)\sqrt{m}t^{n+m-1}(a_0|01\rangle_{34} +$$
$$a_1A_{nm}|10\rangle_{34})|nm-1\rangle_{56}, \quad (C7)$$

$$|\Delta_1(\beta,-\beta_1)\rangle_{123456} = |1,\beta/t\rangle_1|0,-\beta_1/t\rangle_2$$
$$\sum_{n=1}^{\infty}\sum_{m=0}^{\infty} (-1)^n c_{0n}(\alpha)c_{1m}(\alpha_1)\sqrt{n}t^{n+m-1}(a_0|01\rangle_{34} - a_1A_{nm}|10\rangle_{34})|n-1m\rangle_{56} +$$
$$|0,\beta/t\rangle_1|1,-\beta_1/t\rangle_2 \sum_{n=0}^{\infty}\sum_{m=1}^{\infty} (-1)^n c_{0n}(\alpha)c_{1m}(\alpha_1)\sqrt{m}t^{n+m-1}(a_0|01\rangle_{34} -$$
$$a_1A_{nm}|10\rangle_{34})|nm-1\rangle_{56}, \quad (C8)$$

$$|\Delta_2(-\beta,-\beta_1)\rangle_{123456} = |2,-\beta/t\rangle_1|0,-\beta_1/t\rangle_2$$
$$\sum_{n=2}^{\infty}\sum_{m=0}^{\infty} c_{0n}(\alpha)c_{1m}(\alpha_1)\sqrt{\frac{n(n-1)}{2!}}t^{n+m-2}(a_0|01\rangle_{34} + a_1A_{nm}|10\rangle_{34})|n-2m\rangle_{56} +$$

$$|0,-\beta/t\rangle_1|2,-\beta_1/t\rangle_2 \sum_{n=0}^{\infty}\sum_{m=1}^{\infty} c_{0n}(\alpha)c_{1m}(\alpha_1)\sqrt{\frac{m(m-1)}{2!}}t^{n+m-2}(a_0|01\rangle_{34} +$$

$$a_1A_{nm}|10\rangle_{34})|nm-1\rangle_{56} + |1,-\beta/t\rangle_1|1,-\beta_1/t\rangle_2$$
$$\sum_{n=1}^{\infty}\sum_{m=1}^{\infty} c_{0n}(\alpha)c_{1m}(\alpha_1)\sqrt{nm}t^{n+m-2}(a_0|01\rangle_{34} + a_1A_{nm}|10\rangle_{34})|n-1m-1\rangle_{56}, \quad (C9)$$

$$|\Delta_2(\beta,-\beta_1)\rangle_{123456} = |2,-\beta/t\rangle_1|0,-\beta_1/t\rangle_2$$

$$\sum_{n=2}^{\infty}\sum_{m=0}^{\infty}(-1)^n c_{0n}(\alpha)c_{1m}(\alpha_1)\sqrt{\frac{n(n-1)}{2!}}t^{n+m-2}(a_0|01\rangle_{34}-a_1A_{nm}|10\rangle_{34})|n-2m\rangle_{56}+$$
$$|0,-\beta/t\rangle_{1_1}|2,-\beta_1/t\rangle_2\sum_{n=0}^{\infty}\sum_{m=1}^{\infty}(-1)^n c_{0n}(\alpha)c_{1m}(\alpha_1)\sqrt{\frac{m(m-1)}{2!}}t^{n+m-2}(a_0|01\rangle_{34}-a_1A_{nm}|10\rangle_{34})|nm-1\rangle_{56}+|1,-\beta/t\rangle_1|1,-\beta_1/t\rangle_2$$
$$\sum_{n=1}^{\infty}\sum_{m=1}^{\infty}(-1)^n c_{0n}(\alpha)c_{1m}(\alpha_1)\sqrt{nm}\,t^{n+m-2}(a_0|01\rangle_{34}-a_1A_{nm}|10\rangle_{34})|n-1m-1\rangle_{56}.$$
(C10)

Amplitude factor is given by

$$A_{nm}(\alpha,\alpha_1)=\frac{c_{1n}(\alpha)c_{0m}(\alpha_1)}{c_{0n}(\alpha)c_{1m}(\alpha_1)}=\frac{\alpha_1(n-|\alpha|^2)}{\alpha(m-|\alpha_1|^2)}.$$
(C11)

Approximating $t \to 1$, one obtains formula (15).

## Acknowledgement

The work was supported by Act 211 Government of the Russian Federation, contract № 02.A03.21.0011.

**List of figures**

**Figure 1(a,b)**
A schematic representation of generation of hybrid entangled states. (a) Hybrid state (9) composed of the coherent states and single-rail qubit is generated by interaction of the even SCS with the state (7) on HTBS followed by registration of $n$ photons at auxiliary third mode. (b) Extension of the previous case for generation of the hybrid state (16) comprised from the coherent states and dual-rail single photon. The scheme involves additional HTBS for interaction of coherent state $|0,-\beta\rangle_2$ with the original state (14). Commercially achievable avalanche photodiode (APD) being a highly sensitive semiconductor electronic device are used for registration of the measurement outcomes.

**Figure 2(a-c)**
Plot of the success probabilities (a) $P_0^{(b)}$ and (b) $P_1^{(b)}$ to produce the balanced hybrid entangled states in dependency on the displacement amplitude $\alpha$ on which original qubit is displaced on HTBS in indistinguishable manner. The balanced states are generated if (a) vacuum and (b) single photon, respectively, is registered in auxiliary mode. Fidelity $Fid_1(\alpha)$ of the generated states is shown in (c). The plots are made for the following values of transmittance $t=0.8$ (curve 1), $t=0.9$ (curve 2) and $t=0.99$ (curve 3).

**Figure 3(a-c)**
The success probabilities of (a) $P_{00}^{(b)}(\alpha)$ (a) and (b) $P_{00}^{(b)}(\alpha)$ of generation of the balanced hybrid entangled state as function of the displacement parameter $\alpha$. The curves are made for the following values of the transmittance $t=0.8$ (curve 1), $t=0.9$ (curve 2) and $t=0.99$ (curve 3). (c) The success probabilities $P_{01}(\alpha)=P_{10}(\alpha)$ to generate unbalanced hybrid entangled state in dependencies on $\alpha$. The curves with $t=0.9$ and $t=0.99$ almost coincide with each other (curve 2). The curve 1 is for $t=0.8$. Curves 3 is the sum of the success probabilities for all $n+m\leq 5$ in the case of $t=0.99$.

**Figure 4**
The success probability of $P^{(b)}$ to produce hybrid entangled state (30) in dependency on the transmittance $t$. The quantity $P^{(b)}$ is sum of four events $P^{(b)}=P_{00}^{(b)}(\alpha=1/\sqrt{2})+P_{11}^{(b)}(\alpha=1/\sqrt{2})+P_{01}(\alpha=1/\sqrt{2})+P_{10}(\alpha=1/\sqrt{2})$. The curve is obtained for balanced target state (14) with $|a_0|=|a_1|=1/\sqrt{2}$.

*(a)*

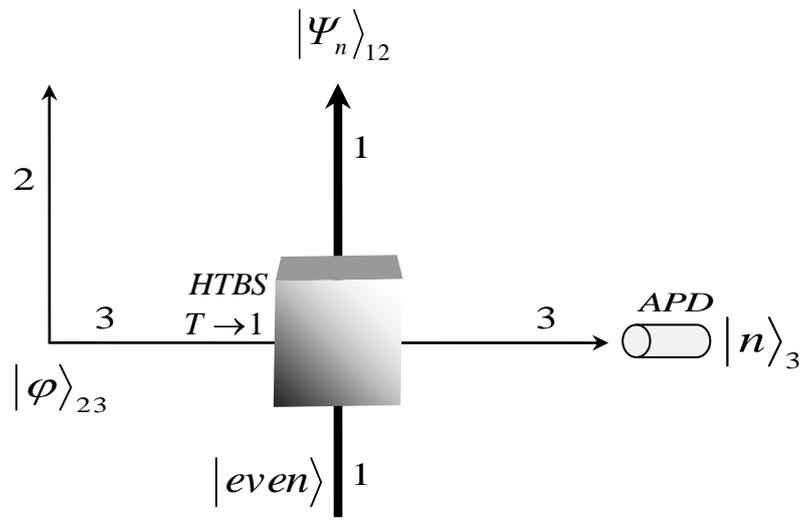

*(b)*

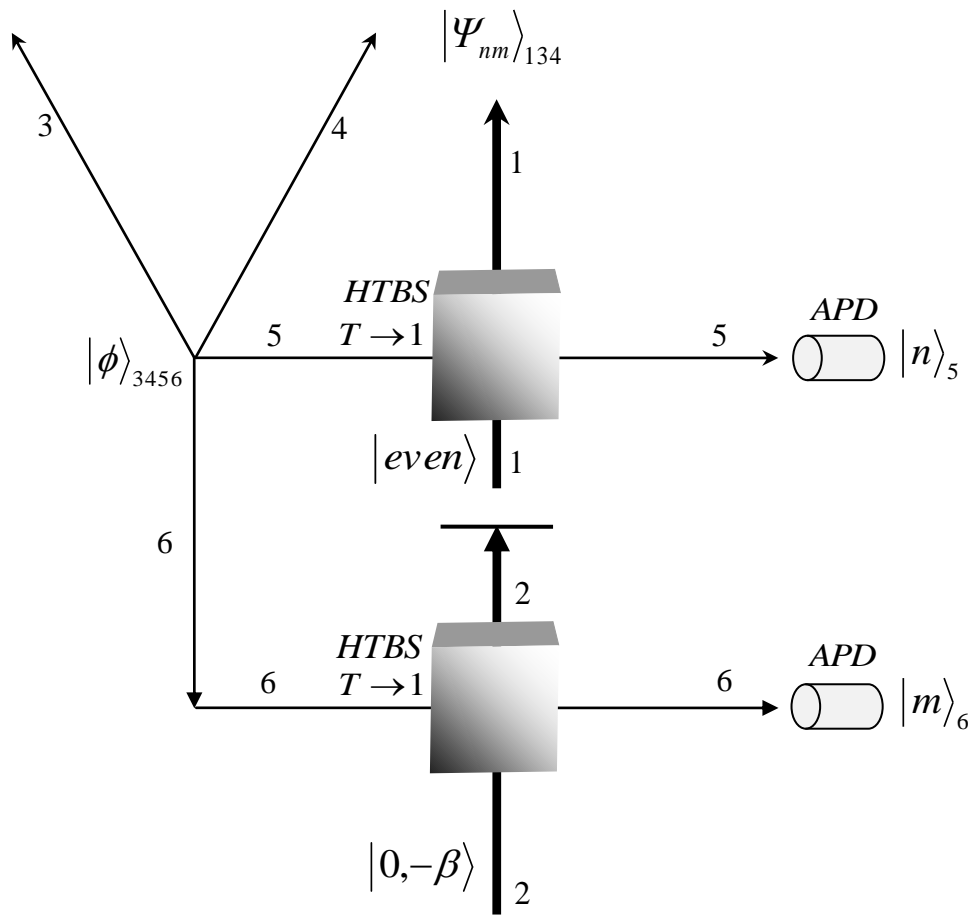

**Figure 1 (a,b)**

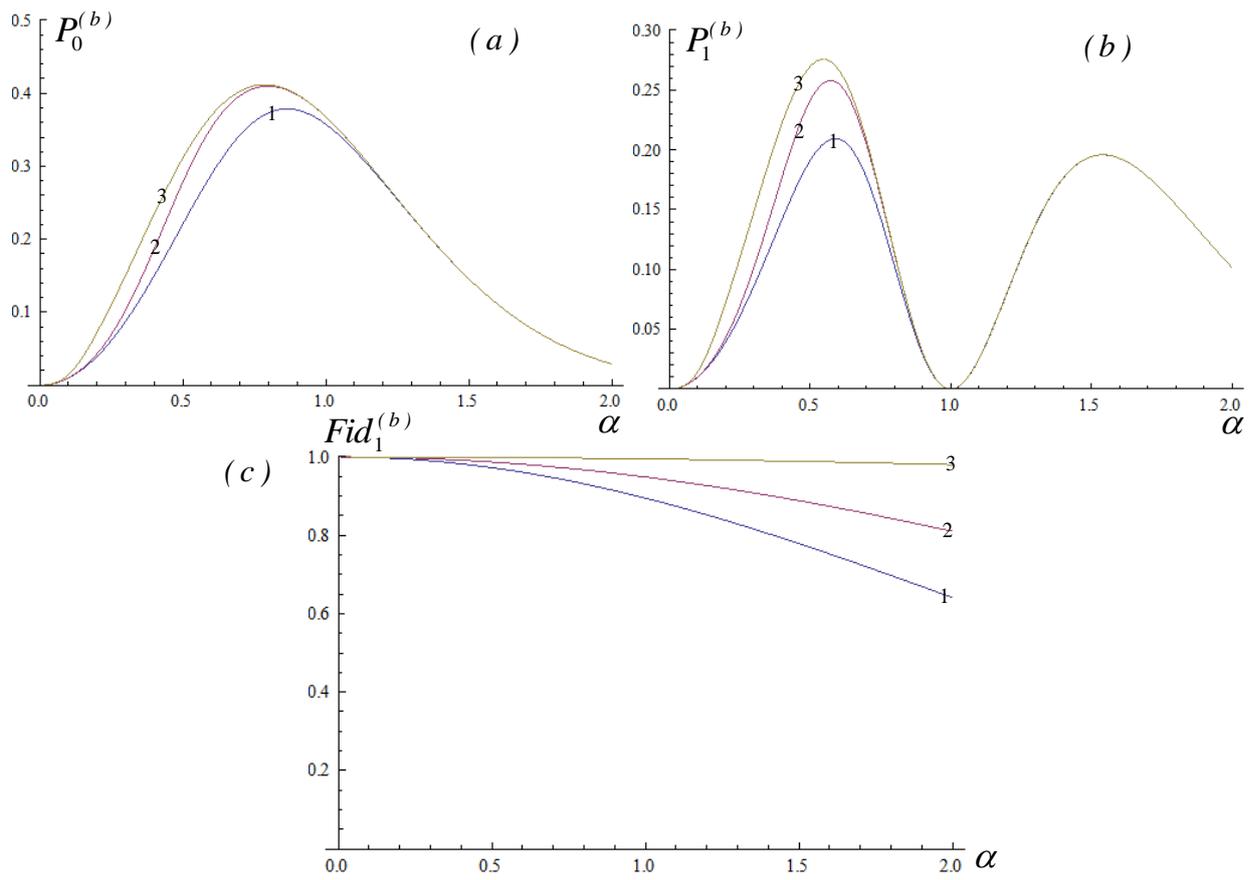

**Figure 2 (a-c)**

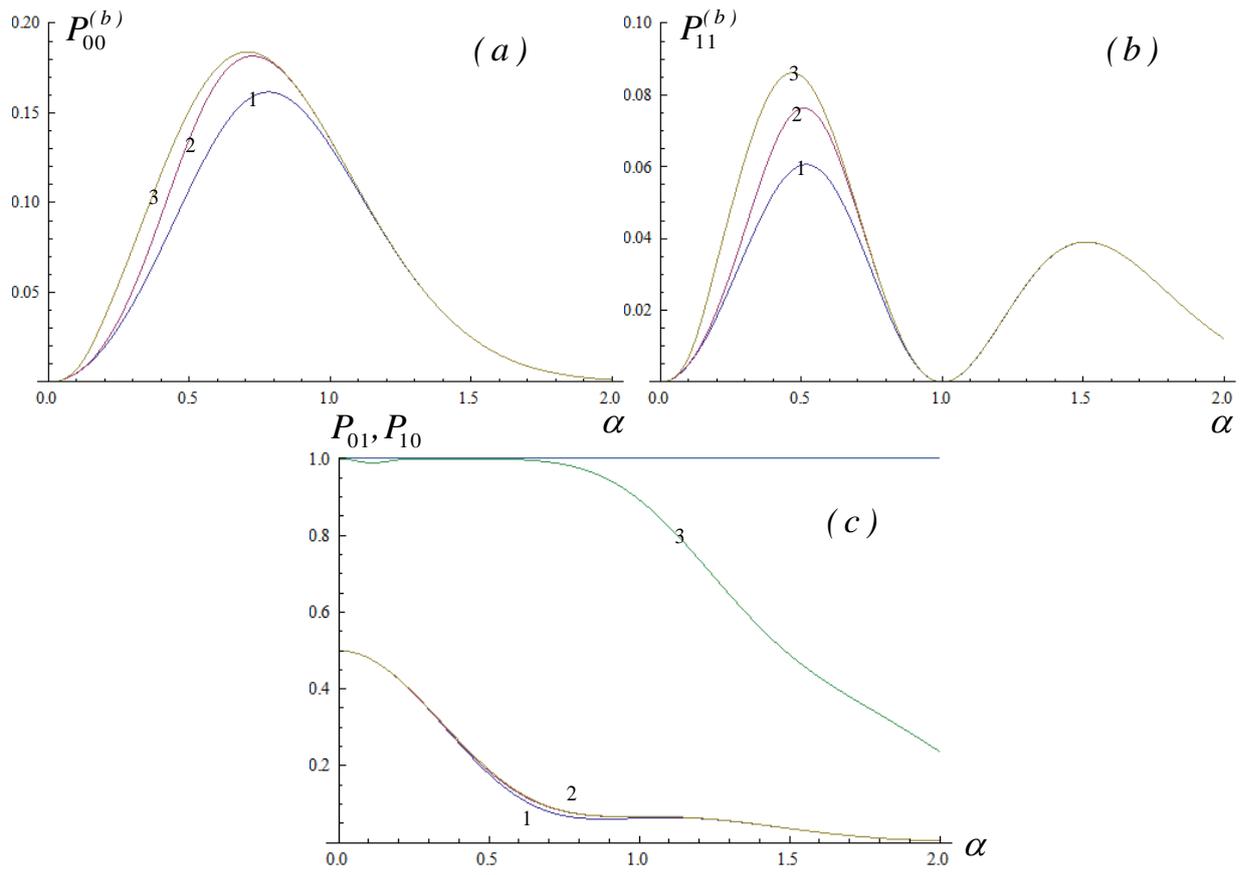

Figure 3 (a-c)

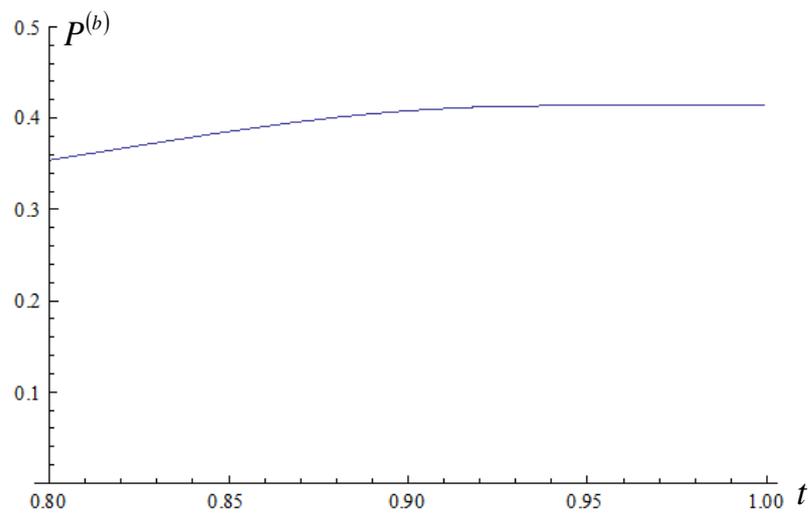

Figure 4